\documentclass[a4,12pt]{article}
\usepackage{bm}
\usepackage{amsmath}
\usepackage{epsfig}\normalsize
\begin{document}
\begin{center}
{\bf \large Constraints  on nuclear matter properties from QCD susceptibilities.}\\[2ex]

M. Ericson$^{1,2}$, G. Chanfray$^1$\\
$^1$ 
  Universit\'e de Lyon, Univ.  Lyon 1, 
 CNRS/IN2P3,\\ IPN Lyon, F-69622 Villeurbanne Cedex\\
$^2$ Theory division, CERN, CH-12111 Geneva 

\begin{abstract}
We establish the interrelation between the QCD scalar response of the nuclear medium and its response to a scalar probe coupled to nucleons, such as the scalar meson responsible for the nuclear binding. The relation that we derive applies at the nucleonic as well as at the nuclear levels. Non trivial consequences follow. In particular  it opens the possibility of relating medium effects in the scalar meson exchange or three-body forces of nuclear physics to QCD lattice studies of the nucleon mass.  
\end{abstract}
Pacs: 24.85.+p 11.30.Rd  12.40.Yx 13.75.Cs 21.30.-x
\end{center}
 \section{Introduction}

The spectrum of scalar-isoscalar excitations is quite different in the vacuum and in the nuclear medium.  In the second case it includes low lying nuclear excitations  and also  two quasi-pion states, {\it i.e.}, pions dressed  by particle hole-excitations. All these lie at lower energies than the vacuum scalar excitations which start at $2\,m_\pi$. We have shown in previous works \cite{CEG03,CE05,CDEM06} that this produces a large increase of the magnitude of the scalar QCD susceptibility over its vacuum value. We have expressed the origin of this increase as arising from the mixing of the nuclear response to a scalar probe coupled to nucleonic scalar density fluctuations into the QCD scalar response.   

It is natural to investigate also the reciprocal problem of the influence of the QCD scalar response to a probe which couples to the quark density fluctuations on the ordinary nuclear scalar response of nuclear physics, which is the object of the present work. We will study this influence not only for what concerns the  nuclear excitations but also for a single nucleon for which only nucleonic excitations are involved. If this influence indeed exists, does it lead to non-trivial observable consequences~? We will show that this is the case, with one main application. It is the possibility to infer medium effects in the propagation of the scalar meson which binds the nucleus from QCD results, such as the lattice ones on the evolution of the nucleon mass with the pion mass.  

Our article is organized as follows. In  section {\bf 2} we illustrate the mixing notion of the nuclear response into the QCD one, and vice-versa
in the framework of a nuclear chiral model with a scalar and vector meson exchange. We show that this mutual influence also exists  at the nucleonic level. In section {\bf 3} we discuss the influence of the quark structure of the nucleon on the scalar response of nuclear physics in a framework which also  incorporates confinement effects.

\section{ Mutual influence of the scalar QCD response and nuclear physics 
response }
\subsection{Study in a nuclear chiral model}
We first remind how the usual nuclear physics response to a scalar field enters in the QCD susceptibility. For this, following ref. \cite{CEG03}, we start from the expression of the  modification of the quark  condensate in the nuclear medium, 
$ \Delta\langle\bar q q\rangle(\rho)=\langle\bar{q}q\rangle (\rho)-\langle\bar{q}q\rangle_{vac}$. We first use, as in ref. \cite{CEG03}, its expression for a collection of independent nucleons~:
\begin{equation}
 \Delta\langle\bar q q\rangle(\rho)=Q_{S}\,\rho_S,
\end{equation}
where $\rho_S$ is the nucleon scalar density.
We have introduced  the scalar charge of the  nucleon, $Q_S =\sigma_N / 2\,m_q$, which represents the scalar number of quarks of the nucleon.
The susceptibility of the nuclear medium, $\chi_S^A$, is the derivative of the quark scalar density with respect to the quark mass. We define it in such a way that it represents a purely nuclear contribution with the vacuum susceptibility substracted off ~:
\begin{equation}
\chi_S^A = \left({\partial \Delta\langle\bar q q(\rho) \over \partial 
m_q}\right)_{\mu}
= \left({\partial (Q_{S}\,\rho_S)\over \partial m_q}\right)_{\mu} . 
\end{equation}
Here the derivatives are taken at constant chemical potential. This expression contains two terms. One arises from the derivative of 
$Q_S$, which by definition is  the free nucleon QCD scalar susceptibility, $\chi_S^N= \partial Q_S/\partial m_q$. The second one involves the 
derivative of the nucleon  density $\rho_S$. This last contribution is itself built of two pieces, one involves antinucleon excitations and is small  \cite{CEG03}. The other one involves, as shown in ref. \cite{CEG03}, the nuclear response $\Pi_0=-2 M_N p_F/\pi^2$. In this case it is the free
Fermi gas one since no interactions between nucleons have been introduced. The result  of this derivation is summarized in the following equation:
\begin{equation}
\chi_S^A = \rho_S\, \chi_N^S \,+ \,2\, Q_S^2\, \Pi_0\,.
\label {CHISGUICHON}
\end{equation}
It says that the nuclear susceptibility is, as expected, the sum of a term arising from the individual nucleon response, {\it {i.e.}}, from the nucleonic excitations, and of a  term linked to the nuclear excitations. This decomposition survives the introduction of the interactions between the nucleons, as will be shown next.
The previous result has been generalized in ref. \cite{CE05} to an assembly of nucleons interacting through a scalar and a vector meson exchanges,  working  at the mean field level as in relativistic mean field theories. The original point with respect to standard relativistic theories is that, 
following our suggestion of ref.  \cite{CEG01}, the nuclear scalar field is identified with a scalar field of the linear sigma model. Rather than the sigma field, chiral parner of the pion, this is a chiral invariant, denoted $S$, associated with the chiral circle radius. 
Nevertheless the nuclear scalar field influences the condensate. Ignoring pion loops the distinction between the two scalar fields, the chiral invariant nuclear one and the non chiral invariant sigma one, wil be ignored. 

In ref. \cite{CE05}, the condensate was  obtained as the derivative of the grand potential with respect to the quark mass (Feynman-Hellmann theorem) and the susceptibility as the derivative of the condensate, both being taken at constant chemical potential. The result for the susceptibility, as given in ref. \cite{CE05}, reads~:
\begin{equation}
\chi_S=\left( {\partial\langle\bar q q\rangle\over\partial m_q}\right)_\mu
\simeq -2\,{\langle\bar q q\rangle_{vac}^2\over 
f_\pi^2}\,\left({\partial\bar S
\over\partial c}\right)_\mu .\label{CHIS}
\end{equation}
$\bar S\equiv f_\pi\,+\,\bar s$   is the expectation value of the chiral invariant scalar field and $c=f_\pi\,m^2_\pi$ is the symmetry breaking parameter of the model used in \cite{CE05}. The quantity $\left({\partial\bar S /\partial c}\right)_\mu$ was shown in ref. \cite{CE05} to be related 
to the in-medium sigma propagator~:
\begin{equation}
\left({\partial\bar S \over\partial c}\right)_\mu= -D^*_\sigma(0)= 
{1\over  m^{*2}_\sigma}\,-\,{g^2_S\over  m^{*2}_\sigma}\, 
\Pi_{S}(0)\,{1\over  m^{*2}_\sigma}
\label{CHISEFF}\end{equation}
where $\Pi_{S}(0)$  is the full RPA scalar polarization propagator and  $m^*_\sigma$ is the in-medium sigma mass,   obtained from the second derivative of the energy density with respect to the order parameter~:
\begin{equation}
m^{*2}_\sigma ={\partial^2 \varepsilon\over\partial\bar s^2}=V''(\bar 
s)\,+\,
{\partial\left(g_S\right)\over \partial \bar s}\,\rho_S
= m^{2}_\sigma\left(1 \,+\,{3\bar s\over f_\pi}\,+\,{3\over 2}\left({\bar 
s\over
f_\pi}\right)^2\right)\,
\label{MSIGMA}
\end{equation}
where the potential $V$ responsible for the spontaneous symmetry breaking is the standard quartic one of the linear sigma model, 
$V=(m^2_{\sigma}/2)\,(s^2+s^3/f_{\pi}+s^4/(4f^2_{\pi}))$. At this stage the nucleons are sructureless and hence we ignore the medium renormalization of $g_S$, {\it i.e.}  we take $g_S$ to be independent of $s$. The mean scalar field $\bar s$  being negative, the term linear in $\bar s$  lowers the 
sigma mass by an appreciable amount ($\simeq 30$ \%  at $\rho_0$). This is the chiral dropping  associated with chiral restoration   \cite{HKS99} and arising from the $3\sigma$ interaction as depicted in fig 1.
  
Since we are interested only in the medium effects the vacuum value of the quantity $\left({\partial\bar S /\partial c}\right)_\mu=1/{m_\sigma}^2$ has to be subtracted off in eq.  (\ref{CHISEFF}) and the purely nuclear suceptibility, $\chi _S^A$,  writes~:
\begin{equation}
\chi _S^A\,=\,{2\,{\langle\bar q q\rangle_{vac}^2\over f_{\pi }^2} }
\left[ {3\,\bar s/ f_{\pi}\,+\,{3\over 2}\left( \bar s/  f_{\pi 
}\right)^2\,\over  m^{*2}_{\sigma }}
\,+\, {g^2_S\over  m^{*2}_{\sigma}} \,  \Pi_{S}(0)\, {1\over  
m^{*2}_{\sigma }}\right] .
\label{CHIAS}
\end{equation}
We see that $ \chi _S^A$ receives two types of contributions, the second denoted as $(\chi _S^A)^{nuclear}$ being proportionnal to the full RPA scalar response  $\Pi_{S }$ (the response to the scalar nuclear field is  $g^2_S\,\Pi_{S }$). The corresponding proportionality factor $r$ between this second contribution and $g^2_S\,\Pi_S$ writes, to leading order, {\it i.e.}, neglecting the medium modification of the sigma mass~:
\begin{equation}
r = \frac{(\chi _S^A)^{nuclear}}{g^2_S\,\Pi_{S }(0)}=2\,{\langle \bar q 
q\rangle _{vac}^2\over  f_{\pi }^2\, m_{\sigma }^4} \simeq 
2\,{( Q^s_S)^2\over g^2_S}
\label{R}
\end{equation}
where we have introduced the nucleon scalar charge $Q_S^s$ from the scalar field,  defined below. In the sigma model the free nucleon sigma commutator is the sum  of two contributions, one  arising from the pion cloud, which depends on the mean value of the squared pion field, {\it i.e.}, on the scalar number of pions in the nucleonic cloud. In the mean field approximation where pion loops are ignored this term does not appear.  The other one, $Q_S^s$, arises from the scalar meson  \cite{B94,DCE96,SKR06}. It is linear in the $\sigma$ field~: 
\begin{equation} 
\label{QSS}
Q_S^s= {\sigma_N^s\over 2 m_q} = -{ \langle\bar q q\rangle_{vac}\over 
f_{\pi }} \int d^3 r\,\langle N| 
\sigma (\vec{r})|N\rangle
 = -{\langle\bar q q\rangle_{vac}\over f_{\pi }}\,{g_S\over m_{\sigma}^2}
\end{equation}
which establishes relation (\ref{R}) if we ignore the in-medium modification of $Q_S^s$, {\it i.e.}, the difference beween $ m^*_{\sigma}$ and $ m_{\sigma}$.

We now turn to the first part of $\chi_S^A$ which depends on the mean scalar field $\bar s$. We will show that it provides an information on the nucleon susceptibility. For this we investigate the low density limit of eq. (\ref{CHIAS}). In this case, $\bar s$ reduces to $\bar s=-g_S\,\rho_S/ m_{\sigma }^2,$ and we can ignore the term in $\bar s^2$ as well as the difference beween $ m^*_{\sigma}$ and $ m_{\sigma}$. In this limit the 
first term in the expression (\ref{CHIAS}) of  $\chi_S^A$ is linear in the density. In the decomposition  of eq. (\ref{CHISGUICHON}) for $\chi_S^A$, it obviously  belongs to the individual nucleon contribution, $\rho_S\,{\chi^N_S},$ to the nuclear susceptibility. Writing the linear term  explicitly in eq. (\ref{CHIAS})  we deduce the free  nucleon scalar susceptibility from the scalar field, 
$(\chi^N_S)^s$~: 
\begin{equation}
(\chi^N_S)^s \,=\,-2\,{\langle\bar q q\rangle_{vac}^2\over f_{\pi 
}^3}\,{3\,g_S
\over m_{\sigma}^4}, 
\label {CHISN}
 \end{equation}
which is negative. The existence of a contribution to the nucleon susceptibility from the scalar field as  given by the expression \ref{CHISN} is a new information provided by this study with interacting nucleons. We have obtained it from the low density expression of $\chi^A_S$. Another way to derive it is  from the derivative with respect to the quark mass of the scalar charge $Q^s_S$ of  eq. (\ref{QSS})~:
\begin{equation}
({\chi^N_S})^s\,= {\partial Q_S^s\over \partial m_{q}} ={\partial \over 
\partial m_q}
\left( -\,{\langle\bar q q\rangle_{vac}\over f_{\pi}}\,{g_S\over 
m_{\sigma}^2}\right). \label{CHISS2}
\end{equation}
Using the fact that, in the model, $\langle\bar q q\rangle_{vac}/ f_{\pi}$ does not depend on $m_q$, only the derivative of the sigma mass with respect to $m_{q}$ enters which, according to the Feynman-Hellmann theorem,  is linked to the sigma commutator, $\sigma_{\sigma}$,  of the $\sigma$. In the linear sigma model the  derivative with respect to the quark mass is replaced by the derivative with  respect to the symmetry breaking parameter, $c=f_\pi\,m_\pi^2$, keeping the other original parameters  of the model, $\lambda$ and $v$, constant. The result is~: 
\begin{equation}
\sigma_{\sigma }= m_q\,{\partial m_{\sigma}\over \partial m_q} ={3\over 
2}\,{m^{2}_{\pi }
\over m_{\sigma}} .
\end{equation}
When inserted in eq. (\ref{CHISS2}), it leads for   $({\chi^N_S})^s$ to the expression of eq. (\ref{CHISN}).

We have seen that the nuclear part of the susceptibility is related to the nuclear response to the scalar field by the relation \ref{R}. Similarly we will  show the nucleonic piece of the susceptibility, $({\chi^N_S})^s$, is related  to the scattering amplitude of the scalar meson on the nucleon.
Indeed, in  the expression(\ref{MSIGMA}) of $m^{*2}_{\sigma }$  there is a  term linear in density which is obtained from the low density expression~: $3\,\bar s \,m^{*2}_{\sigma }\simeq-(3\,g_S/ f_{\pi} )\,\rho_{S}$. This term represents an optical  potential  for the scalar meson  
propagation. The corresponding $\sigma N$  scattering amplitude, $T_{\sigma N}$, which can also be evaluated directly from the graph of  fig. 1, is equal to~:
\begin{equation}
T_{\sigma N}= -3\,g_S / f_{\pi }\label{TSN}.
\end{equation}
\begin{figure}
\begin{center}
\epsfig{file=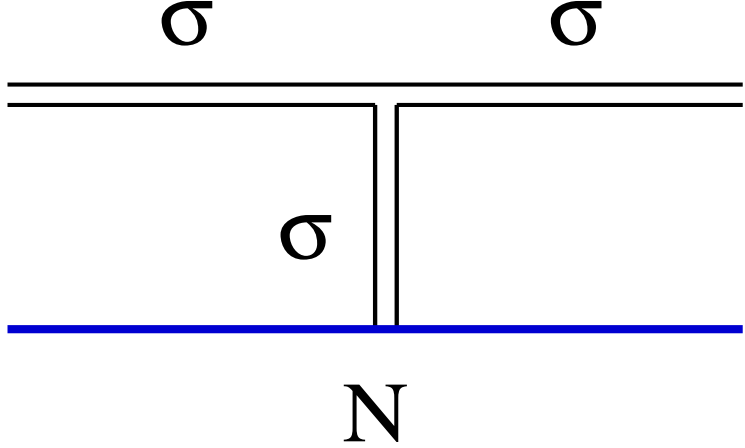,width=10.0cm,height=6.0cm,angle=0}
\end{center}
\caption{Contribution to the sigma-nucleon scattering amplitude responsible for the lowering of the sigma mass in the medium.}
\label{}      
\end{figure} 
We are now in a situation to relate the nucleon scalar susceptibility (eq. (\ref {CHISN})) to the sigma-nucleon amplitude of eq. (\ref{TSN}), with the result~:
\begin{equation}
r'=\frac{(\chi_S^N)^s}{T_{\sigma N}}=\frac{2\,(Q_S^s)^2}{g_S^2} .
\label{CHISNS}
 \end{equation}
We observe that the proportionality factor between   $T_{\sigma N}$ and 
$(\chi_S^N)^s$ which is
 $2\,(Q_S^s)^2 / g_S^2$, is identical to the one which involves  the purely nuclear excitations. The quantity $g_S$ which appears in the present factor  is due to the $\sigma NN$ coupling constant. Adding now the two effects from the nucleonic and nuclear excitations the 
total QCD scalar susceptibility of the nuclear medium (vacuum value substracted) can therefore be related to the total response, $T^A$, to the nuclear scalar field through~:
\begin{equation}
\chi _S^A = {2\,(Q_S^s)^2\over g_S^2} T^A\label{RESP}
\end{equation}
where the two members include both the individual nucleon contribution and the one arising from the nuclear excitations,with~:
\begin {equation}
T^A= \rho_{S} \,T_{\sigma N}\, +\, g^2_S\, \Pi_{S} .
\end {equation}
The last term on the r.h.s. represents the  influence of the Born part of the $\sigma N$ amplitude (in-medium corrected in particular for the Pauli effect) while the first piece arises from  the non-Born part linked to  nucleonic excitations.

\section{Connection with lattice data}

Since we have introduced QCD quantities such as the QCD scalar response, it is now interesting to connect our results to lattice simulations  of the evolution of the nucleon mass with the pion mass, equivalently the quark mass.  At present they do not cover the physical region but only the 
region  beyond $m_{\pi } \simeq 400\,MeV$ and in order to reach the physical nucleon mass an extraplolation has to be performed. The pion cloud contribution to the nucleon self.energy has a non-analytic behavior in the quark mass, preventing a polynomial expansion in this quantity. For that reason Thomas et al  \cite{TGLY04}  have separated out this contribution. This is done in a model dependent way with different  cut-off forms for the pion loops (gaussian, dipole, monopole) with an adjustable  parameter $\Lambda$. They expand the remaining part in terms of $m^2_{\pi}$  as follows:
\begin {equation}
M_N(m^{2}_{\pi}) = 
a_{0}\,+\,a_{2}\,m^{2}_{\pi}\, 
+a_{4}\,m^{4}_{\pi}\,+\,\Sigma_{\pi}(m_{\pi}, \Lambda).
\label{EXPANSION}
\end{equation}
The best fit value of the parameter $a_{4}$  which fixes the  susceptibility shows little sensitivity to the shape of the form  factor, with a value 
$a_4 \simeq-\,0.5\, GeV^{-3}$, while $a_2 \simeq 1.5\,GeV^{-1}$ (in  a previous work \cite{TLY04} smaller values of $a_2$ and $a_4$ were given~: $a_2\simeq 1\,GeV^{-1}$ and $a_4 \simeq-\,0.23\, GeV^{-3}$). From the expansion of eq. (\ref{EXPANSION})  we can therefore infer 
the non-pionic pieces of the sigma commutator and of the susceptibility~:
\begin{equation}
\sigma_N^{non-pion} = m^2_{\pi} \,{\partial M\over \partial m^2_{\pi }}
~=~a_2\, m^2_{\pi}\, +\, 2\,a_4\,  m^4_{\pi}~\simeq 29\,MeV \,.
\label{SIGMALATTICE}
\end{equation}
It is largely dominated by the $a_2$ term. The corresponding value for $a_2\simeq 1\,GeV^{-1}$ is $\sigma_N^{non-pion}=20\,MeV$. In turn the nucleon susceptibility is~:
\begin{equation}
\chi_S^{ N, non- pion}~= 2\,{\langle\bar q q\rangle_{vac}^2 \over 
f^4_{\pi} }
{\partial \over \partial  m^2_{\pi }}\left({\sigma_N^{non-pion} \over 
m^2_{\pi }}\right)~=  
{\langle\bar q q\rangle_{vac}^2 \over f^4_{\pi} }\,4\,a_{4}
 ~\simeq -5.4\, GeV^{-1} 
\label{CHILATTICE}
\end{equation}
The non-pionic susceptibility is found with a negative sign, as expected from the scalar meson term. In ref. \cite{TGLY04} however, the negative sign is interpreted differently. It is attributed to possible deviations from the  Gellman-Oakes-Renner (GOR) relation which links quark and pion masses. 
Here instead we assume the validity of the GOR relation.

It is then interesting to test if the empirical values from the lattice are compatible with our previous linear sigma model results. We thus tentatively make the following identifications~:
\begin{equation}
Q_S^s~= -{\langle\bar q q\rangle_{vac}\over f_{\pi }}\,{g_S\over 
m_{\sigma}^2}~=
 {\sigma_N^{non-pion}\over (2\, m_q)}\simeq -{\langle\bar q 
q\rangle_{vac}\over f_{\pi }}a_2 = - 2.4,
 \label{QSSLAT}
\end{equation}
with $2\,m_q=12 \,MeV$ (taking $a_2\simeq 1\,GeV^{-1}$ one would get $Q_S^s=1.66$). It is interesting to translate this number into the value of the mean scalar field in the nuclear medium which, to leading order in density, is~:
\begin{equation}
-\bar s= {g_s\,\rho_S\over m^2_\sigma}={Q_S^s\,f_{\pi 
}\,\rho_S\over\langle\bar q q\rangle_{vac}}
={\sigma_N^{non-pion}\over (2\, m_q)}\,{f_{\pi }\,\rho_S\over\langle\bar q 
q\rangle_{vac}}
={a_2\,+\,a_4\,m_\pi^2\over f_\pi}\,\rho_S~.
\end{equation}
At normal density the value is $|\bar s(\rho_0)|\simeq 21\, MeV$, compatible with nuclear phenomenology. The second identification concerns the 
susceptibility. Identifying the value of the linear sigma model with the lattice one, we should have~: 
\begin{equation}
 (\chi_S^N)^{non- pion}~=~-{2 \,(Q_S^s)^2\over g_S^2}\,{3\,g_S\over 
f_{\pi}} 
 =~{\langle\bar q q\rangle_{vac}^2 \over f^4_{\pi} }\,4\,a_{4} 
\end{equation}
which, using the relation (\ref{QSSLAT}) between $a_2$ and $Q_S^s$ leads to~:
\begin{equation}
-a_4={3\over2} \,{(\sigma_N^{non-pion})^2 \over g_S\,f_\pi\,{m_\pi}^4} 
\simeq {3\over 2}{a^2_2 \over g_Sf_{\pi}} =
{3\over g_S f_{\pi}} a^2_2 = 3.5\, GeV^{-3},
\end{equation}
much larger than the lattice value, $-a_4= 0.5 \, {\rm GeV}^{-3}$. For  $a_2\simeq 1\,GeV^{-1}$ one gets $-a_4\simeq 1.2\, GeV^{-3}$, also larger  
than the corresponding lattice value $(0.23 \, GeV^{-3})$.  Thus the linear $\sigma $ model leads to a too large absolute   value of the nucleon scalar  
susceptibility. We remind at this stage that it also fails in another respect, concerning the saturation properties of nuclear matter. The $3\sigma$ coupling present in this model, which lowers the sigma mass in the medium, prevents saturation to occur and produces the collapse \cite{KM74}. 
Said differently the sigma nucleon scattering amplitude, $T_{\sigma N}$,  of the model is too attractive. In fact what we have shown in this work is 
that the two problems are linked since we have found that  $T_{\sigma N}$ and $(\chi _{S}^N)^s$ are related. These two failures are coherent. Some  mechanism  must be at work to  introduce in both  a suppression effect. In  a previous work \cite{CE05} we have invoked the quark meson  coupling model (QMC) \cite{G88,GSRT96} and confinement as a source of cancellation. Indeed,  for three valence quarks confined in a bag of radius $R$,  
Guichon \cite{G05} derived $(\chi_S^{N})^{bag}\simeq\,+\,0.25\, R \simeq 1\,GeV^{-1}$, (for  $R=0.8\,fm$).  Contrary  to the other components which are negative (of paramagnetic nature), it has a positive sign (of the diamagnetic type, linked to quark-antiquark excitations).  In ref (\cite{CE05}) we have introduced phenomenologically in  the nucleon mass evolution of the linear sigma model a parameter, $\kappa _{NS}$ which embodies the scalar response of the nucleon from confinement:
\begin {equation}
M_N^* = M_N + g_S \bar s +{1\over 2} \kappa_{NS}\bar s^2
\label {MSTAR}
\end {equation}
It allows a proper description of the saturation properties on nuclear matter.It is then natural to extend the linear sigma model description so as to incorporate  effects arising from confinement.  

\section{ Illustration in a hybrid model of the nucleon}
We now want  to generalize our previous results so as to incorporate the confinement aspect. In the following we will introduce a model of the 
nucleon proposed in  ref. \cite{ST99} which is intermediate between the two extreme pictures~: the bag one and the Nambu-Jona-Lasinio (NJL) one which generates a linear sigma model. We will study both the scalar susceptibility of the nucleon and the scattering amplitude of the scalar field on the nucleon and their relation. In this framework we retain  two concepts that were  contained in our previous approach in the linear sigma model~:
({\it i}) the nuclear scalar field is identified with the chiral  field associated with the quark condensate  and  ({\it ii}) part of the nucleon mass originates from this  condensate. This model   consists in the following.  Three constituant quarks moving  in a non-perturbative vacuum are kept together by a central force which mimicks confinement and the effect of the color string tension. The mass $M$ of the constituants quarks originates from the chiral condensate as in the NJL model. The nucleon mass is not $3M$ but, because of the confining force, becomes  $3E(M)$ where the 
M dependence is fixed by the type of force. For illustration we take for simplicity a harmonic force of the form: $((K/4)(1+\gamma_0)\,r^2$, which leads to analytical results. With this particular  potential the nucleon mass is~:
\begin{equation}
M_N =3\,E=3\left( M+{3\over2}\sqrt{K\over E+M} \right) \,.
\end{equation}
It is increased as compared to the value, $3M$, for three independent constituant quarks. Although oversimplified the model  gives an intuitive picture  of the role played by confinement. Since we  assume that the nuclear scalar field is related to the quark condensate, the presence of the mean scalar field in the medium which modifies the condensate with respect to its vacuum value also affects the mass M. The derivative, 
$\partial M/ \partial {\bar s}$, has a non-vanishing value, given by the NJL model,  
\begin{equation}
 {\partial M\over \partial {\bar s}}= g_q = \frac{M}{f_{\pi}}.
\end{equation}
 The nucleon scalar charge, $Q_S$, writes~:  
\begin{equation}
 Q_S ={3\over 2}\, {\partial E\over \partial m_q}  
={3\over 2} \,{\partial E\over \partial M}  {\partial M\over \partial m_q}
\end{equation}
with :
\begin{equation}
 {\partial E\over \partial M}=c_S
={E+3M\over3E+M}\,.
\end{equation}
As $E>M$, $c_S<1$, the nucleon scalar charge is reduced as compared to a collection of three independent  quarks. The nucleon scalar susceptibility, $\chi^N_{S}$, given by the next derivative, is composed of two terms arising respectively from the derivative of $c_S$ and from that of  $\partial M/ \partial m_q$~:
\begin{equation}
\chi^N_{S} = 
 {\partial Q_S\over \partial m_q}= {3\over 2} 
\left[ {\partial c_S\over \partial M}\left({\partial M\over \partial m_q}\right)^2+ c_S \, 
{\partial^2  M\over \partial^2  m_q^2} \right] \\ \nonumber
\label{DOUBLE}
\end{equation}
with:
\begin{equation}
 {\partial c_S\over \partial M}={24 \,(E^2- M^2)\over {(3E+M)}^3}\,.
\end{equation}
Notice that this last derivative is positive since $E>M$ and that it vanishes in the absence of confining force, when $E=M$. Therefore the first part of the expression of $\chi^N_{S}$ represents the part of the susceptibility originating in confinement and, as in QMC, it is positive. We find in the susceptibility written in eq. (\ref{DOUBLE}) the double aspect of the mass, part arising from the constituant quark mass, i.e., from the condensate and part from confinement. The expression  (\ref{DOUBLE}) can descibe the two extreme situations. In the MIT bag model the confined quarks are the current ones, $ M=m_q$, the second term of the susceptibility disappears, only the confinement part  enters. In the NJL model instead  where the 
constituant quarks are unconfined, $E=M,  c_S=1$ and only the second term survives.

For what concerns the  coupling constant, $g_S$~, of the nucleon to the scalar field, it  is given by the derivative of the nucleon mass with respect to the mean scalar field $\bar s$:
\begin{equation} 
g_S = 3\,{\partial E\over \partial \bar s}= 3\,  {\partial E\over \partial 
M}{{\partial M\over \partial \bar s}}
= 3 \,c_S\, g_{q }.
\end{equation}
The nucleon response to the scalar field originating in confinement, $\kappa _{NS}$, follows from the eq. ({\ref {MSTAR}) as the second derivative of the nucleon mass with respect to the scalar field~: 
\begin{equation} 
\kappa _{NS} = 3\,
{\partial^2 E\over \partial \bar s^2}= 3\, 
 {\partial  c_S\over \partial M}\left(
{\partial M\over \partial \bar s}\right)^2\,.
\end{equation}
The ratio, $r_m$,  between the part of the nucleon scalar susceptibility  due to confinement and $\kappa_{NS}$ is
\begin{equation}
r_m={1\over 2}\,{({\partial M\over \partial m_q})^2\over ({\partial M\over 
\partial s})^2}
= {2 \,Q_S^2 \over g_S^2},
\end{equation}
the same ratio, $r$,  as was previously found in the linear sigma model. 

While the quantity  $\kappa _{NS}$ represents the effect of the nucleon internal quark structure, there is another component of the $\sigma N$ amplitude, which is the tadpole term , $T_{\sigma N}$, an effect of the mexican hat chiral potential. For each constituant quark the tadpole amplitude is $t_{\sigma N} = -3\,g_q/f_{\pi}$.  As the scalar number of constituant quarks is $3\,c_S$, the tadpole amplitude for the nucleon writes  ${T_{\sigma N}}^{tadpole} = -3\,g_S/f_{\pi}$, the same expression as in the linear sigma model. This quantity should be compared to the other component  of the susceptibility, ${3\over2} c_S \, {\partial^2  M\over \partial^2  m_q^2}$ . We define $r'_m$ as the corresponding ratio through~:
\begin{equation}
{3\over 2}\, c_S\, {\partial^2  M\over \partial m_q^2} = 
r'_m\,\left(-{3\,g_S\over f_{\pi}}\right)\,. 
\end{equation}
In the semi-bosonized version of the NJL model we have~:
\begin{equation}
{\partial  M\over \partial  m_q}  =-{2 \,{g_q\,\langle\bar q 
q\rangle_{vac}
\over f_{\pi}\,m^2_{\sigma}}} 
\end{equation}
and  
\begin{equation}
{\partial^2  M\over \partial  m_q^2} = -{2 \,g_q\,{\langle\bar q 
q\rangle_{vac}}^2 \over 
{f_{\pi}^3}\,m^4_{\sigma}}
\end{equation}
 in such a way that the ratio $r'_m$ becomes~: 
\begin{equation}
r'_m= {2\, \langle\bar q q\rangle_{vac}^2 \over 
{f_{\pi}^2}\,m^4_{\sigma}} = {2\,Q^2_S\over g^2_S} { \equiv r_m}.
\end{equation}
Since the same ratio applies to the two parts, $r'_m\equiv r_m$, it can be factorized when we add the two pieces of the susceptibility so as
to obtain the relation 
\begin{equation}	
\chi^{N}_{S}=r_m\,\kappa_{NS}\,+\,r'_m\,\left(-\frac{3\,g_S}{f\pi}\right)={2\,Q^2_S\over 
g^2_S}\,T^{total}_N \label{TRUTH}
\end{equation}
which thus also holds in the presence of confinement. Adding the nuclear excitations contribution (term in $\Pi_S$) on both sides of the above 
equation once multiplied by the density we recover the relation (\ref{CHIAS}) between the nuclear values $\chi^{A}_{S}$ and ${T}_A$.

Numerically we have chosen a value of the ratio $E/M \simeq 2.1$, which   leads to a reasonable value for $g_A$ and gives $c_S \simeq 0.7$. It results in a value of the dimensionless parameter $C=(f_{\pi}/ 2 g_S)\,\kappa_{NS} \simeq 0.1$, while the value needed to account for the saturation properties is $C\simeq 1$ \cite{CE07}. In fact the nuclear phenomenology requires a strong suppression of the tadpole term in the $\sigma N$ amplitude (total cancellation occurs for $C=1.5$). On the other hand the lattice results also require a nearly total cancellation of the nucleon scalar susceptibility from the scalar meson by the effect of confinement. For us the two cancellations have a unique origin and description since the total susceptibility and the total $\sigma N $amplitude are related. The condition for a total cancellation between the 
two components of $\chi_S^N$, as approximately required by the phenomenology, writes
\begin{equation}
 {\partial c_S\over \partial M}-{c_S\over 2g_qf_{\pi}} = 0,
\end{equation} 
which should approximately hold for the physical value of M, but  is not fulfilled with our particular form of  $c_S$ where the second part dominates. Even if our particular  model fails to account for the numerical value of $C$ it has the merit to confirm the validity of the relation  between the QCD response and the one to the nuclear scalar field in a situation where confinement enters. The  relation (\ref{TRUTH}) is indeed general and  does not depend on the particular form of $E(M)$.

\section{Applications to nuclear physics}
We can now turn to the quantitative applications of the relation  (\ref{TRUTH}) between $T_N^{total}$ and the scalar susceptibility. The last quantity is known from the lattice expansion. On the other hand the scalar charge which enters the relation  (\ref{TRUTH}) is also determined by this expansion. Therefore the only model dependent quantity to determine the amplitude, $T_N^{total}$, from the lattice expansion is  $g_S$ but this is only a moderate uncertainty. The  important point is that the resulting value of $T^{total}_N$  is  small.

The resulting medium effects in the  propagation of the nuclear scalar field can be written, using the eq. (\ref{TRUTH}) and (\ref{CHILATTICE}), as~:
\begin{equation}
-D_s^{-1}= 
m_\sigma ^2\,+\,{g_S^2\over2\, Q_S^{2}}\,\chi^N_{S}\,\rho_{S} 
\approx m_{\sigma }^2\,+\,g_S^2\,{2\, a_4\over a_2^2}\, \rho _S;
\end{equation}
Numerically, at normal nuclear density, and for a value of the coupling constant $g_S=10 $, the second term on the rhs of the second equation takes the value $0.06 \,GeV^2$ (a similar value is found for the other set of parameters $a_2$ and $a_4$). For a sigma mass of $m_\sigma=0.75\, GeV$,  this represents at $\rho_0$ only a 6\% decrease of the mass, much less that the chiral dropping alone and  in much better agreement with the nuclear phenomenology \cite{CE05,CE07}. 

At this stage, conceptual questions naturally arise. As the lattice  parameter $a_4$ is very small one may conclude that QCD effects related to the nucleon quark substructure and chiral symmetry are simply not visible in nuclear physics as they annihilate each other. This conclusion is however erroneous. They strongly show up at the level of the three-body forces, playing an important role in the saturation mechanism, as explained below. 
For that purpose we introduce a new scalar field $u=s+(\kappa_{NS}/2g_S)\,s^2$ in such a way that the expression of the in-medium nucleon effective 
mass reduces to a simpler form according to~: 
\begin{equation}
M^*_N(\bar u)=M_N\,+\,g_S\,{\bar s}\,+\,{1\over 2}\,\kappa_{NS}\,
\bar s^2 \equiv M_N \,+ \,g_S\, \bar u.
\end{equation}
Expressed in term of the $u$ field, the chiral mexican hat potential takes the form~:
\begin{equation}
V^{chiral}=  V={ m_\sigma^2\over 2} \left(s^2+{s^3\over f_\pi} +{s^4\over 
4\,f^2_\pi}\right)=
 {m_\sigma^2\over 2} \left(u^2 +{ u^3\over f_{\pi}}(1-2C) + {u^4\over 
4}(1-8C+20C^2)\right).
\end{equation}
In the formulation with the $u$ field the three body forces are  contained in the $u^3$ term~:
 \begin{equation}
V^{three-body}=  {m_\sigma^2\over 2} \,{\bar u^3\over f_{\pi}}\,(1\,-\,2C). 
\label{V3}
\end{equation} 
We remind the definition of $C=(\kappa_{NS}f_{\pi})/(2g_S)$. As $\bar u<0$, this force is repulsive for $C>1/2$, which is actually the case. Without confinement, {\it i.e.}, $C=0$, the chiral potential alone  leads to attractive 3-body forces. The important point is that the balance between the effects of the chiral potential and of confinement are not the same in the propagation of the scalar field and in the three body forces. In the first case the amplitude $T^{total}_N$ which governs the sigma self-energy is 
$T^{total}_N= 3\,g_S/f_{\pi} + \kappa_{N S} = (3g_S/f_{\pi})\, (1-2 C/3)$, while in the three body forces  the combination  is $1-2 C$. With $C$  of the order unity, a strong cancellation occurs in $T^{total}_N$ while there is an overcompensation  in the three body potential which becomes repulsive. The existence of repulsive three body forces in relativistic theories is strongly supported by the nuclear phenomenology  \cite{LKR97,GMST06}.  Using the eq. (\ref{CHILATTICE}),(\ref {QSSLAT}) and (\ref{TRUTH}), we can express $C$ in terms of the lattice parameters~:
\begin{equation}
C = {3\over 2}-
 \frac{g_S\,f_{\pi}}{a^{2}_{2}} \,a_4\ \simeq 1.3,
\end{equation} 
which leads to
\begin{equation}
V^{three-body}=  
 -{m_\sigma^2\over 2} \,{\bar u^3\over 
f_{\pi}}\,\left(\frac{g_S\,f_{\pi}}{a^{2}_{2}} \,a_4\,+\,2\right)
\end{equation} 
Notice that  since $a_4$ is small the term $2$ dominates the parenthesis on the r.h.s. The equation of motion gives $\bar u\simeq -g_S\,\rho_S/m^{2}_{\sigma}$. 

Numerically,  for the phenomenological value  $C=1$,   the contribution of the three-body forces to the energy per nucleon is~:
\begin{equation}
\left(\frac{E}{A}\right)^{three-body}	
=\frac{V^{three-body}}{\rho} 20\,\left(\frac{\rho}{\rho_{0}}\right)^2_, MeV. 
\end{equation}
With the lattice value. $C=1.3$, the result is $\simeq 50\%$ larger. 

In the QMC approach of ref. \cite {G88,GSRT96}, the chiral aspect is not considered and hence the higher order terms  in the mexican hat potential are absent.  The three-body potential originates only from confinement through the nucleon scalar response~: 
\begin{equation}
V^{three-body}=  {m_\sigma^2\over 2} \,{\bar u^3\over f_{\pi}}\,(-\,2C^{QMC}). 
\label{V3G}
\end{equation} 
Comparing the expressions (\ref{V3}) and (\ref{V3G}), one sees that, numerically, our phenomenological value $C\simeq 1$ is equivalent  
to  $C^{QMC} = 0.5 $, which is close to the actual value of the QMC model.
 
\section{Conclusion}
In summary we have studied in this work the interplay between the two nuclear responses to  probes which couple either to nucleon or to quark scalar density fluctuations. We have found that the two responses are reflected in each other. The scaling coefficient involves the nucleon scalar charge.  Both responses incorporate the individual nucleon contributions to the nuclear response in such a way that our result holds not only at  the level of the nuclear excitations but also for the nucleonic ones. The  response of a nucleon to the nuclear scalar field is linked to its QCD scalar susceptibility. We have first established this results in the linear sigma model, adopting the view that the scalar field is the chiral invariant scalar field of this model, in which case it is linked to the quark condensate.

However the linear sigma model has serious shortcomings, first in nuclear physics where it makes nuclear matter collapse  instead of saturate.  In QCD as well it fails to account for the value of the nucleon scalar susceptibility, for which it predicts  too large a magnitude as compared 
to the lattice result. In our views the two problems are not distinct but they are automatically linked and we have  attributed them to the absence of confinement in the description. In a second step we have improved  our approach to incorporate this effect.  For the nucleon we have adopted a hybrid image of three constituant quarks sitting in a non perturbative vacuum  and kept together by a confining potential. The nucleon mass thus originates in part from the quark condensate and in part from confinement. We have retained the concept that the nuclear scalar field has a 
relation to the scalar field associated with the  chiral quark condensate. In this situation the presence of the nuclear scalar mean field affects 
the condensate and  hence the constituant quark mass. We have shown  that, in this model, the relation between  the scalar meson 
self-energy and the nucleon QCD scalar susceptibility remains the same as in the linear sigma model.

The existence of relations between  nuclear physics parameters (such as the opticel potential for the propagation of thr scalar field, the three-body potential)  and those of QCD  opens the possibility of a description of the properties of nuclear matter using as inputs the parameters of the lattice expansion of QCD. We have adopted this approach in ref.{(\cite {CE07}}) and it has been  successful. With parameters close to those provided by the lattice 
expansion we have been able to reproduce the saturation properties of nuclear matter. This coherence, which  it is not a priori acquired, suggests the
validity of such an approach. It supports the idea that  a  part of the nucleon mass originates in the quark condensate and that the nuclear scalar field plays a role in the restoration  of chiral 
symmetry in nuclei. We have found, both in the lattice expansion results  and in the nuclear phenomenology, the need for a strong cancellation of the 
chiral effects by confinement in the sigma propagation. It follows that the sigma mass remains stable in the medium.  Confinement nevertheless 
shows up very neatly in the three body potential where it dominates the attractive  
chiral effects, giving rise to repulsive three-body forces.


\end{document}